\documentclass[conference]{IEEEtran}
\IEEEoverridecommandlockouts
\usepackage{cite}
\usepackage{amsmath,amssymb,amsfonts}
\usepackage{algorithmic}
\usepackage{graphicx}
\usepackage{textcomp}
\usepackage{xcolor}
\usepackage{cite}
\usepackage{amsmath,amssymb,amsfonts}
\usepackage{algorithmic}
\usepackage{graphicx}
\usepackage{textcomp}
\usepackage{xcolor}
\usepackage{nicefrac}
\usepackage{graphics}
\usepackage{float}
\usepackage{graphicx}
\usepackage{amsmath}
\usepackage{color}
\usepackage{dsfont}
\usepackage{bm}
\usepackage{upgreek}
\usepackage{arydshln}
\usepackage{enumerate}
\usepackage{amsthm}
\usepackage{graphicx}
\usepackage{threeparttable}
\usepackage{caption}
\usepackage{multirow}
\usepackage{color}
\usepackage{xfrac}
\usepackage{array}
\usepackage{nomencl}
\usepackage{hyphenat}
\usepackage{cite}
\usepackage{amsmath,amssymb,amsfonts}
\usepackage{algorithmic}
\usepackage{graphicx}
\usepackage{textcomp}
\usepackage{xcolor}
\usepackage{url}
\def\BibTeX{{\rm B\kern-.05em{\sc i\kern-.025em b}\kern-.08em
		T\kern-.1667em\lower.7ex\hbox{E}\kern-.125emX}}

\newcommand{\be}{\begin{equation}}
\newcommand{\ee}{\end{equation}}

\newcommand{\norm}[1]{ || #1 ||}
\newcommand{\trace}[1]{ \mathrm{tr}( #1 )}
\newcommand{\mb}[1]{\mathbf{#1}}
\newcommand{\bs}[1]{\boldsymbol{#1}}

\newcommand{\virg}[1]{\textquotedblleft#1\textquotedblright}

\newcommand{\kronVtinvmTTy}{\widehat{\mb{V}}_{1,Ty}^{-T/2}\otimes\widehat{\mb{V}}_{1,Ty}^{-1/2}}

\newcommand{\cvec}[1]{ \mathrm{vec}\left(  #1 \right) }

\newcommand{\ovec}[1]{\underline{\mathrm{vec}}(#1)}

\newcommand{\tonde}[1]{\left( #1 \right)  }
\newcommand{\quadre}[1]{\left[  #1 \right]  }
\newcommand{\graffe}[1]{\left\lbrace   #1 \right\rbrace   }

\def\BibTeX{{\rm B\kern-.05em{\sc i\kern-.025em b}\kern-.08em
    T\kern-.1667em\lower.7ex\hbox{E}\kern-.125emX}}
\begin{document}

\title{Robust Semiparametric DOA Estimation in non-Gaussian Environment\\
\thanks{The work of S. Fortunati, A. Renaux and F. Pascal has been partially supported by DGA under grant ANR-17-ASTR-0015.}
}

\author{\IEEEauthorblockN{Stefano~Fortunati, Alexandre~Renaux, Fr\'{e}d\'{e}ric~Pascal}
	\IEEEauthorblockA{\textit{Universit\'{e} Paris-Saclay, CNRS, CentraleSupel\'{e}c, Laboratoire des signaux et syst\`{e}mes,} \\
		91190, Gif-sur-Yvette, France. \\
		e-mails: \{stefano.fortunati, alexandre.renaux, frederic.pascal\}@centralesupelec.fr}
}

\maketitle

\begin{abstract}
A general non-Gaussian semiparametric model is adopted to characterize the measurement vectors, i.e.\ the \textit{snapshots}, collected by a linear array. Moreover, the recently derived \textit{robust semiparametric efficient} $R$-estimator of the data covariance matrix is exploited to implement an original version of the MUSIC estimator. The efficiency of the resulting $R$-MUSIC algorithm is investigated by comparing its Mean Squared Error (MSE) in the estimation of the source spatial frequencies with the relevant Semiparametric Stochastic Cram\'{e}r-Rao Bound (SSCRB).
\end{abstract}


\section{Introduction}
In array processing, the word \virg{\textit{robustness}} has been declined in many different ways depending on the specific application at hand. According to that branch of mathematical statistics started with the seminal works of Huber and Hampel \cite{huber_book,hampel_book}, in this paper we will focus on the \textit{distributional robustness} of inference procedures. Distributionally robust methodologies are, in fact, of fundamental importance in situations where a not perfect \textit{a priori} knowledge of the statistical behaviours of the collected measurements (summarized in their joint distribution) leads to a significant degradation in the performance from the expected nominal one. Since the exact input data distribution is rarely \textit{a priori} known, the design of robust approaches has been gaining considerable attention in many applicative fields \cite{Zoubir,book_zoubir}.

In this paper, we deal with the Direction of Arrival (DOA) estimation of $K$ sources from a set of $L$ independent, identically distributed (i.i.d.) measurement vectors $\{\mb{z}_l\}_{l=1}^L$, also called \textit{snapshots}, collected by a linear sensor array. As it can be observed from the vast literature on this topic, a lot of effort has been devoted to derive optimal DOA estimation algorithms to be used in a specific (and a priori known) signal and disturbance environment that, for the sake of tractability, is generally assumed to be Gaussian. This simplifying assumption, however, violates the everyday practice that highlights the non-Gaussian, heavy-tailed behaviour of the data. 

As suggested in \cite{visuri}, \cite{Esa_DOA} and \cite{abeida_SP_paper} a suitable class of non-Gaussian, heavy-tailed distributions able to describe the statistical behaviour of the snapshots is given by the Complex Elliptically Symmetric (CES) distributions. The CES model generalizes all the commonly used array processing data models (Gaussian and Compound Gaussian ones, among others) and its reliability has been validated by extensive measurement campaigns and related data analysis (see \cite{Greco_data_analysis,Esa} and references therein). Formally, a snapshot $\mathbb{C}^N \ni \mb{z}_l \sim CES_N(\mb{z};\bs{\theta},h)$ is said to be \textit{zero-mean} CES-distributed if its probability density function (pdf) can be expressed as \cite{Esa}:
\be
\label{true_CES}
\begin{split}
	p_{Z}(\mb{z}_l|\bs{\theta}, h) 
	=|\bs{\Sigma}(\bs{\theta})|^{-1} h \left(  \mb{z}_l^{H}
	\bs{\Sigma}(\bs{\theta})^{-1}\mb{z}_l \right),
\end{split}
\ee     
where $\bs{\theta} \in \Theta$ is a \textit{finite-dimensional} vector, containing the parameters of interest, that parametrizes the covariance/scatter matrix $\bs{\Sigma}(\bs{\theta})$.\footnote{Note that this formulation includes the case where the whole scatter matrix $\bs{\Sigma}$ is unknown. In fact, we can always define $\bs{\theta} \triangleq \mathrm{vec}(\bs{\Sigma})$, where the vectorization operator $\mathrm{vec}$ is formally defined in the notation section.} The other \textit{infinite-dimensional} parameter, characterizing the actual CES data distribution, is the \textit{density generator} $h$ that belongs to the set $\mathcal{G} \triangleq \graffe{ h: \mathbb{R}^{+} \rightarrow \mathbb{R}^{+} \left|   \int_{0}^{\infty}t^{N-1}h(t)dt < \infty, \int p_Z =1 \right. }$ \cite{Esa}.

Since, as said before, in practical application it is unrealistic to assume the \textit{a priori} knowledge of the specific data distribution, the density generator $h$ has to be considered as a nuisance function. Consequently, the joint pdf of a set of $L$ i.i.d.\ zero-mean, CES-distributed snapshots has to be considered as an element of a \textit{semiparametric} model of the form \cite{For_SCRB,For_SCRB_complex}:
\be
\mathcal{P}_{\bs{\theta},h} \triangleq \graffe{\prod\nolimits_{l=1}^{L}p_Z(\mb{z}_l;\bs{\theta},h); \bs{\theta} \in \Theta, h \in \mathcal{G}  }.
\ee 
Therefore, the estimation of $\bs{\theta} \in \Theta$ has to be framed in the context of semiparametric inference problems and it has to be handled by means of distributionally robust algorithms that does not rely on the knowledge of the density generator $h$.

The goal of this paper is then to investigate the performance of a subspace-based MUSIC DOA estimator built around a new \textit{distributionally robust} and  \textit{semiparametric efficient} $R$-estimator of the snapshot covariance matrix. Such original $R$-estimator, whose exploitation in radar signal processing is still at its infancy, has been firstly proposed by Hallin, Oja and Paindaveine in \cite{Hallin_Annals_Stat_2} for Real ES data, while its extension to CES data is provided in our recent work \cite{Sem_eff_est_TSP}.

Before moving forward, it is important to underline that other semiparametric estimators for the covariance matrix of a set of non-Gaussian data vectors have already been proposed in signal processing literature. As en example, we refer to \cite{Hugo, PASCAL_el} for an \textit{Empirical Likelihood} approach to structured covariance matrix estimation. 

The paper is organized as follows. At first, a general semiparametric CES snapshot model is provided in Sec. \ref{sec_CES_snap_model}. This model generalizes and encompasses as special case the Gaussian snapshot model commonly adopted in array processing literature. Then, Sec. \ref{sec_R_MUSIC} presents the main ideas behind the distributionally robust and semiparametric efficient $R$-estimator of the snapshot covariance matrix along with its exploitation to derive a MUSIC-based DOA estimation algorithm. In order to assess the semiparametric efficiency of the derived $R$-MUSIC estimator, we compare its Mean Squared Error (MSE) with the Semiparametric Stochastic Cram\'er-Rao Bound (SSCRB) recently derived in \cite{For_SCRB_complex,For_EUSIPCO_19} and recalled here in Sec. \ref{sec_SSCRB}. The simulation results about the MSE performance of the proposed $R$-MUSIC estimator are provided in Sec. \ref{numerical} while our concluding remarks are collected in Sec. \ref{conclus}. 

\textit{Notation}: In the rest of the paper, italics indicates scalar quantities ($a$), lower case and upper case boldface indicate column vectors ($\mathbf{a}$) and matrices ($\mathbf{A}$), respectively. The $(i,j)$ entry of a matrix $\mb{A}$ is indicated as $a_{ij}\triangleq [\mb{A}]_{i,j}$. A matrix $\mb{A}$ whose first top-left entry is constrained to be equal to 1, i.e. $a_{11} \triangleq 1$, is indicated as $\mb{A}_1$. The operator $\mathrm{vec}$ maps column-wise the entry of an $N \times N$ matrix $\mathbf{A}$ in an $N^2$-dimensional column vector $\cvec{\mb{A}}$. The operator $\ovec{\mb{A}}$ defines the $N^2-1$-dimensional vector obtained from $\cvec{\mb{A}}$ by deleting its first element, i.e.\ $\cvec{\mb{A}} \triangleq [a_{11},\ovec{\mb{A}}^T]^T$. The subscript \virg{0} indicates the actual (or \textit{true}) quantities characterizing the data generating process. Specifically, $\bs{\theta}_0$, $h_0$ and $p_0(\mb{z}_l)\triangleq p_Z(\mb{z}_l|\bs{\theta}_0, h_0)$ defines the true parameter vector, the true density generator and the true data pdf, respectively. The symbol $=_d$ stands for \virg{has the same distribution as}. Finally, let us define the matrices $\mb{P} \triangleq \quadre{\mb{e}_2|\mb{e}_3|\cdots| \mb{e}_{N^2}}^T$, where $\mb{e}_i$ is the $i$-th vector of the canonical basis of $\mathbb{R}^{N^2}$, and $\Pi^{\perp}_{\cvec{\mb{I}_N}}=\mb{I}_{N^2} - N^{-1}\mathrm{vec}(\mb{I}_N)\mathrm{vec}(\mb{I}_N)^T$.

\section{The semiparametric CES snapshot model}\label{sec_CES_snap_model}
The aim of this Section is to the introduce the non-Gaussian snapshot model that we are going to assume for the measurement collected by the sensor array. At first, we briefly recall the basic properties of CES distributions. Then, the CES snapshot model is presented and its advantages with respect to the classical Gaussian data model are discussed and analysed.

\subsection{Basics on CES distributions}\label{subsec_CES}
The theory of CES distributions has been extensively discussed in many dedicated works, both in statistics and signal processing literature. Among the many, we refer to the excellent tutorial paper \cite{Esa} and to our previous works \cite{For_SCRB, For_SCRB_complex} where CES distributions have been framed in the context of semiparametric models. Here, for the sake of clarity, we limit ourselves to provide a very short summary that may help the non-expert reader to go through the next sections of this paper.  

Let $\mathbb{C}^{N \times N} \ni\bs{\Sigma}_0 \triangleq \bs{\Sigma}(\bs{\theta}_0)$ be the true scatter matrix, assumed to be of full-rank, parametrized by $\bs{\theta}_0 \in \Theta$. Then, any zero-mean, CES-distributed vector $\mb{z}_l$ can be written as
\be\label{st_rep}
\mb{z}_l =_d \sqrt{\mathcal{Q}}\bs{\Sigma}_0^{1/2}\mb{u}_l,\quad \mathrm{where:}
\ee
\begin{itemize}
	\item[\textit{i})] $\mb{u}_l \sim \mathcal{U}(\mathbb{C}S^N)$ is a complex random vector uniformly distributed on the complex unit $N$-sphere,
	\item[\textit{ii})] $\mathcal{Q}=_d \mb{z}_l^H\bs{\Sigma}_0^{-1}\mb{z}_l \triangleq Q_l \sim p_{\mathcal{Q},0}(q) = \nicefrac{\pi^{N}}{\Gamma(N)} q^{N-1} h_0 (q)$ is called \textit{2nd-order modular variate},
	\item[\textit{iii})] The covariance matrix of $\mb{z}_l$ can be expressed as function of the scatter matrix $\bs{\Sigma}_0$ and $\mathcal{Q}$ as $\mb{M}_0 \triangleq E\{\mb{z}_l\mb{z}_l^H\} = N^{-1}E\{\mathcal{Q}\}\bs{\Sigma}_0$.
\end{itemize}
For identifiability reason as well as for ease of problem interpretation, we impose the equality between the covariance matrix $\mb{M}_0$ and the scatter matrix $\bs{\Sigma}_0$. To this end, as direct consequence of point \textit{iii}), we need to constrain the density generator $h_0$ to belong to the following (constrained) set of functions:
\be\label{const}
\bar{\mathcal{G}} \triangleq \graffe{h \in \mathcal{G} |E\{\mathcal{Q}\} = N},
\ee    
where $E\{\mathcal{Q}\} \triangleq \int_{0}^{+\infty}q p_{\mathcal{Q},0}(q) dq = \frac{\pi^{N}}{\Gamma(N)}\int_{0}^{+\infty}q^{N} h_0 (q)dq$. The expectation operator with respect to pdfs depending on the \virg{constrained} density generator $h_0 \in \bar{\mathcal{G}}$ will be indicated as $\bar{E}\{\cdot\}$. Finally, we define as \textit{shape matrix} $\mb{V}_{0}$ a normalized version of the covariance/scatter matrix. Common normalizations are $\trace{\mb{V}_{0}}=N$ or $|\mb{V}_{0}|=1$. Here, for the reasons discussed in \cite{Sem_eff_est_TSP}, we choose the following $\mb{V}_{1,0}\triangleq \bs{\Sigma}_0/[\bs{\Sigma}_0]_{1,1}$ leading to shape matrices having the first top-left element equal to 1.

\subsection{CES distributed snapshots}
Let us first recall the classical Gaussian snapshot model widely used in array processing literature. We consider an array of arbitrary geometry with $N$ omnidirectional sensors. Moreover, let us assume to have $K$ narrowband sources impinging on the array from $K$ different spatial frequencies collected in the vector $\bs{\nu}=(\nu_1,\ldots,\nu_K)^T$. The \textit{steering matrix} $\mb{A}(\bs{\nu})\triangleq [\mb{a}(\nu_1)|\cdots|\mb{a}(\nu_K)] \in \mathbb{C}^{N \times K}$ is defined as the matrix whose $k^{\mathrm{th}}$ column is given by the \textit{steering vector} $\mb{a}(\nu_k)$ characterizing the array geometry. 

The classical snapshot model is given by \cite{Stoica_CRB_2, Stoica_CRB}:
\be\label{cla_model}
\mb{A}_0\mb{s}_l + \mb{n}_l \triangleq \mb{z}_l  \sim \mathcal{CN}(\mb{0},\bs{\Sigma}(\bs{\theta}_0)),
\ee
\begin{itemize}
	\item $\mb{A}_0$ indicated the steering matrix evaluated at the true spatial frequency vector $\bs{\nu}_0$, i.e. $\mb{A}_0 \triangleq \mb{A}(\bs{\nu}_0)$,
	\item $\mathbb{C}^K \ni \mb{s}_l \sim \mathcal{CN}(\mb{0},\bs{\Gamma}_0), \forall l$ is the Gaussian source vector and $\bs{\Gamma}_0$ is the source correlation matrix,
	\item $\mathbb{C}^N \ni \mb{n}_l \sim \mathcal{CN}(\mb{0},\sigma^2_0\mb{I}_N), \forall l$ is the white Gaussian measurement noise with power $\sigma^2_0$.
	\item The unknown parameters can be collected in 
	\be\label{theta}
	\bs{\theta}_0 \triangleq [\bs{\nu}_0^T,\bs{\zeta}_0^T,\sigma_0^2]^T \in \Theta,
	\ee 
	where $\Theta \subseteq [-0.5,0.5)^K \times \mathbb{R}^{K^2} \times \mathbb{R}^+$ and $\bs{\zeta}_0$ is a vector set up by the $K$ diagonal entries as well as the $K(K+1)-2K$ real and imaginary parts of the off-diagonal entries of the source correlation matrix $\bs{\Gamma}_0$.
\end{itemize}
From the model in \eqref{cla_model}, the snapshot covariance/scatter matrix can easily be obtained as:
\be\label{cov_struc}
\bs{\Sigma}(\bs{\theta}_0) \triangleq \bar{E}\{\mb{z}_l\mb{z}_l^H\} = \mb{A}_0\bs{\Gamma}_0\mb{A}_0^H + \sigma^2_0\mb{I}_N.
\ee
Note that the covariance structure in \eqref{cov_struc} does not necessary require the sources or the noise to be Gaussian distributed, but it holds true for any uncorrelated random vectors $\mb{s}_l$ and $\mb{n}_l$ having finite second-order moments.
  
Even if widely used due to its analytical tractability, the linear snapshot model in \eqref{cla_model} suffers from a strong limitation: the source and noise vectors are assumed to be two independent and Gaussian-distributed vectors, and consequently their linear combination (i.e.\ the resulting snapshot) is also Gaussian-distributed. In practical applications however, this Gaussianity assumption is not a realistic one, at least for the noise/clutter contribution. In addition, when wrongly adopted, it can lead to severe performance degradation. For this reason, instead of assuming two separate Gaussian models, one for the sources and one for the disturbance, we prefer to adopt directly a non-Gaussian model for the snapshots without specifying the source and disturbance models \cite{visuri}, \cite{Esa_DOA}, \cite{abeida_SP_paper}. In particular, we assume that the $L$ i.i.d. snapshots are distributed as \cite{For_SCRB_complex, For_EUSIPCO_19}:
\be\label{CES_model}
\mb{z}_l  \sim CES_N(\mb{z},\bs{\theta}_0, h_0),
\ee      
where $\bs{\theta}_0 \in \Theta$ is the parameter vector defined in \eqref{theta}, while $h_0 \in \bar{\mathcal{G}}$ as to be considered as an additional \textit{nuisance} function. The snapshot model in \eqref{CES_model} has two big advantages with respect to the classical one:
\begin{itemize}
	\item[\textit{i})] It is able to catch the non-Gaussian, heavy-tailed behaviour of the collected measurements,
	\item[\textit{ii})] It is \textit{semiparametric} in nature, i.e. it can encompass a wide range of non-Gaussian models without any need of pre-selecting a specific data distribution. In fact, the actual density generator $h_0$ is considered as an infinite-dimensional unknown parameter.
\end{itemize}

If, on one hand, the generality of the semiparametric CES model in \eqref{CES_model} guarantees that the risk of model misspecification is minimized \cite{SPM}, on the other hand its semiparametric nature asks for more sophisticated, \textit{distributionally robust}, inference procedure to estimate the parameters of interest, i.e. the source spatial frequency vector $\bs{\nu}_0$ contained in $\bs{\theta}$. 

To conclude, it is worth noticing that the semiparametric CES snapshot model in \eqref{CES_model} encompasses, as special case, the classical Gaussian one. In fact, the model in \eqref{cla_model} can be obtained from the one in \eqref{CES_model} by setting $h_0 = \exp(-t)$ as density generator for the snapshots.

\section{The $R$-MUSIC algorithm}\label{sec_R_MUSIC}
Given the measurement model in \eqref{CES_model}, a major goal of any array processing inference is to estimate the spatial frequency vector $\bs{\nu}_0$ (i.e. the DOA) from the set of collected snapshots $\{\mb{z}_l\}_{l=1}^L$. Among the variety of DOA estimation algorithms proposed in the array processing literature, we will focus here on that subclass described by a (vector-valued) function $\bs{\Delta} \in \mathcal{H}$ characterized by the following two properties:
\begin{itemize}
	\item[P1] $\bs{\Delta}$ depends on the snapshots $\{\mb{z}_l\}_{l=1}^L$ only through their estimated covariance matrix $\widehat{\bs{\Sigma}}$:
	\be
	\label{h_1}
	\mathcal{H} \ni \bs{\Delta}: \widehat{\bs{\Sigma}} \mapsto \bs{\nu} = (\nu_1,\ldots,\nu_K)^T
	\ee
	\item[P2] $\bs{\Delta}$ is a positively homogeneous function of order zero:
	\be
	\bs{\Delta}(a \widehat{\bs{\Sigma}}) = \bs{\Delta}(\widehat{\bs{\Sigma}}),\; \forall a>0.
	\ee
\end{itemize}
It is easy to verify that the class $\mathcal{H}$ encompasses all the subspace-based methods, and in particular, the MUSIC algorithm as special cases. In particular, let $\mb{V}_{1,0}\triangleq \bs{\Sigma}_0/[\bs{\Sigma}_0]_{1,1}$ be the shape matrix as defined in subsec.\ \ref{subsec_CES}. Moreover, let $\widehat{\mb{V}}_1$ be an estimate of $\mb{V}_{1,0}$ obtained from the collected snapshots $\{\mb{z}_l\}_{l=1}^L$. Then the MUSIC \textit{pseudospectrum} is given by \cite{MUSIC}:
\be\label{MUSIC_PS}
P_M(\nu) \triangleq \quadre{\sum\nolimits_{n=K+1}^N |\mb{a}(\nu)^H\hat{\mb{r}}_n|^2}^{-1},
\ee  
where, as before, $\nu$ is the variable representing the spatial frequency, $\mb{a}(\nu)$ is the steering vector and $K$ is the total number of sources (assumed to be \textit{a priori} known). The vectors $\{\hat{\mb{r}}_{K+1},\ldots,\hat{\mb{r}}_{N}\}$ forms an orthogonal basis for the so called \textit{noise subspace} and they can be obtained as the $N-K$ eigenvectors corresponding to the $N-K$ smallest eigenvalues of the estimated shape matrix $\widehat{\mb{V}}_1$. The MUSIC estimator $\bs{\Delta}_M$ of the source spatial frequencies can be obtained by searching for the position of the first $K$ local maxima of $P_M(\nu)$:
\be
\label{h_M}
\bs{\Delta}_M: \widehat{\mb{V}}_1 \mapsto \hat{\bs{\nu}} = \underset{\bs{\nu}}{\mathrm{argmax}}\; P_M(\nu).
\ee
As we can see from \eqref{MUSIC_PS}, the calculation of the MUSIC pseudospectrum does not rely on the \textit{a priori} knowledge of the joint distribution of the snapshots, so it can be considered as a proper semiparametric DOA estimation algorithm. However, its distributional robustness and (semiparametric) efficiency strongly depends on the choice of the shape matrix estimator. In \cite{For_EUSIPCO_19} we showed that the commonly adopted robust $M$-estimators, such as Tyler's or Huber's estimators, does not lead to semiparametric efficient DOA estimates. Therefore, in this paper, we propose the use of an original $R$-estimator $\widehat{\mb{V}}_{1,R}$ that has been proved to possess two desirable properties when applied to CES data: the distributional robustness and the semiparametric efficiency.    

\subsection{An original $R$-estimator of the shape matrix}
The shape matrix estimator that we are going to introduce belongs to the class of $R$-estimators. This name is motivated by the fact that this family of robust estimators rely on the properties of the \textit{ranks} \footnote{Let $\mathcal{V} \triangleq \{Q_l\}_{l=1}^L$ be a set of continuous random variables and let $\mathcal{V}_o \triangleq \{Q_{L(1)}<Q_{L(2)}<\ldots<Q_{L(L)}\}$ be its relevant ordered (in an ascending way) set. Then, the rank $r_l$ of $Q_l \in \mathcal{V}$ is its position index in $\mathcal{V}_o$.} of a set of order statistics. The in-depth theoretical analysis of this $R$-estimator can be found in \cite{Sem_eff_est_TSP}, while here we report only the final expression along with a short discussion about its practical implementation. 

To define the $R$-estimator $\widehat{\mb{V}}_{1,R}$, we need a \textit{preliminary estimator} of the shape matrix. Even if any $\sqrt{L}$-consistent estimators will do, it is advisable to use the Tyler's estimator  $\widehat{\mb{V}}_{1,Ty}$ due to its robustness properties \cite{Tyler1}. In particular, let  $\{\mb{z}_l\}_{l=1}^L$ a set of CES-distributed snapshots as in \eqref{CES_model}, then $\widehat{\mb{V}}_{1,Ty}$ can be obtained as the convergence point ($p \rightarrow \infty$) of the following iteration \cite{Pascal1,Pascal2}:
\be
\label{Ty_schape}
\left\lbrace  \begin{array}{l}
	\widehat{\bs{\Sigma}}^{(p+1)} =\frac{N}{L}\sum_{l=1}^{L}\nicefrac{\mb{z}_l\mb{z}_l^H}{\mb{z}_l^H[{\bs{\Sigma}}^{(p)}]^{-1}\mb{z}_l}
		\\
	\widehat{\mb{V}}_{1,Ty}^{(p+1)} \triangleq \nicefrac{\widehat{\bs{\Sigma}}^{(p+1)}}{[\widehat{\bs{\Sigma}}^{(p+1)}]_{1,1}}.
\end{array}\right. 
\ee 
Consequently, by relying on $\widehat{\mb{V}}_{1,Ty}$ in \eqref{Ty_schape} and according to \cite{Sem_eff_est_TSP}, the $R$-estimator of the (complex-valued) shape matrix $\mb{V}_{1,0}$ can be expressed as:\footnote{Related Matlab code can be found at \url{https://github.com/StefanoFor/Robust-semiparametric-efficient-R-estimator-for-shape-matrices}.}
\be
\label{com_one_step_R}
\begin{split}
	\ovec{\widehat{\mb{V}}_{1,R}} & = \ovec{\widehat{\mb{V}}_{1,Ty}} -\frac{1}{L\hat{\alpha}}\quadre{\mb{L}_{\widehat{\mb{V}}_{1,Ty}} \mb{L}_{\widehat{\mb{V}}_{1,Ty}}^H}^{-1} \\
	\times &\mb{L}_{\widehat{\mb{V}}_{1,Ty}}\sum_{l=1}^{L}K_{\mathrm{vdW}}\tonde{\frac{r_l^\star}{L+1}} \mathrm{vec}(\hat{\mb{u}}^\star_l(\hat{\mb{u}}^\star_l)^H),
\end{split}
\ee
\begin{itemize}
	\item $\mb{L}_{\widehat{\mb{V}}_{1,Ty}} \triangleq \mb{P} \tonde{\kronVtinvmTTy} \Pi^{\perp}_{\cvec{\mb{I}_N}}$, and $\Pi^{\perp}_{\cvec{\mb{I}_N}}$ and $\mb{P}$ are defined in the notation section,
	\item $\{r_l^\star\}_{l=1}^L$ are the ranks of the continuous random variables $\hat{Q}^\star_l \triangleq \mb{z}_l^H[\widehat{\mb{V}}_{1,Ty}]^{-1}\mb{z}_l$, $l=1,\ldots,L$,
	\item $\hat{\mb{u}}^\star_l \triangleq \nicefrac{[\widehat{\mb{V}}_{1,Ty}]^{-1/2}\mb{z}_l}{\sqrt{\hat{Q}^\star_l}}$, $l=1,\ldots,L$,
	\item The data-dependent term $\hat{\alpha}$ is given in \cite[Sec. V.B]{Sem_eff_est_TSP},
	\item $K_{\mathrm{vdW}}(u) \triangleq - \Phi_G^{-1}(u)$
	where $\Phi_G^{-1}$ indicates the inverse function of the cdf of a Gamma-distributed random variable with parameters $(N,1)$.
\end{itemize}

As amply discussed in \cite{Sem_eff_est_TSP}, the advantage of the $R$-estimator in \eqref{com_one_step_R} with respect to other robust competitors is in the fact that $\widehat{\mb{V}}_{1,R}$ is distributionally robust as the Tyler's or Huber's $M$-estimators but, unlike them, it is also \textit{semiparametric efficient}. Roughly speaking, this means that $\widehat{\mb{V}}_{1,R}$ achieves the Semiparametric Cram\'{e}r-Rao bound (SCRB) \cite{For_SCRB, For_SCRB_complex} on the estimation of the shape matrix $\mb{V}_{1,0}$ irrespective of the unknown density generator $h_0$.   
In Sec. \ref{numerical}, we will investigate whether the semiparametric efficiency of $\widehat{\mb{V}}_{1,R}$ in the estimation of $\mb{V}_{1,0}$ will lead to a better performance of the MUSIC DOA estimator based on it. Specifically, the semiparametric efficiency of the MUSIC functional $\hat{\bs{\nu}} = \bs{\Delta}_M(\widehat{\mb{V}}_{1,R})$ given in Sec. \ref{sec_R_MUSIC} will be assessed by comparing its MSE with the relevant Semiparametric Stochastic CRB (SSCRB). 

\section{The Semiparametric Stochastic CRB}\label{sec_SSCRB}
A classical result in array processing is the so-called \textit{Stochastic Cram\'er-Rao Bound} \cite{Stoica_CRB_2,MUSIC,Stoica_CRB}. In particular, given the Gaussian snapshot model in \eqref{cla_model}, the Stochastic CRB provides us with a lower bound on the MSE of any (unbiased) estimators of the spatial frequency vector $\bs{\nu}_0$ in the presence of two \textit{finite-dimensional} nuisance terms, i.e.\ the source correlation vector $\bs{\zeta}_0$ and the noise power $\sigma_0^2$. Clearly, this result is no longer valid if the semiparametric CES model in \eqref{CES_model} is assumed for the collected snapshots. However, a generalization of the Stochastic CRB for estimation problem based on \eqref{CES_model} has been recently derived in \cite{For_SCRB_complex}. Specifically, in \cite{For_SCRB_complex}, a lower bound to the MSE of any $\sqrt{L}$-consistent, distributionally robust, estimators of $\bs{\nu}_0$ has been obtained in the presence of the (finite-dimensional) nuisance terms $\bs{\zeta}_0$ and $\sigma_0^2$ along with the \textit{infinite-dimensional} nuisance density generator $h_0$. Such Semiparametric Stochastic CRB can be expressed as:
\be
\label{SSCRB}
\mathrm{SSCRB}(\bs{\nu}_0|\bs{\zeta}_0,\sigma_0^2, h_0) = \frac{N(N+1)\sigma_0^2}{2L\bar{E}\{\mathcal{Q}^2\psi_0(\mathcal{Q})^2\}} \mb{C}(\bs{\nu}_0,\bs{\zeta}_0)^{-1},
\ee  
where $\mathcal{Q}$ is defined in subsec. \ref{subsec_CES}, the function $\psi_0$ is such that $\psi_0(t) \triangleq d \ln h_0(t)/dt$ and $\mb{C}(\bs{\nu}_0,\bs{\zeta}_0)$ is given by:
\be
\label{C_mat}
\mb{C}(\bs{\nu}_0,\bs{\zeta}_0) \triangleq \mathrm{Re}\tonde{\mb{D}_0^H \Pi^{\perp}_{\mb{A}_0} \mb{D}_0}\odot \tonde{\bs{\Gamma}_0\mb{A}_0^H\mb{\Sigma}_0^{-1}\mb{A}_0\mb{\Gamma}_0}^T,
\ee
where $\odot$ is the Hadamard product, $\mb{D}_0 \triangleq \quadre{\mb{d}_{0,1},\cdots,\mb{d}_{0,K}}$, $\mb{d}_{0,k}\triangleq \left. d\mb{a}(\nu_k)/d\nu_k \right|_{\nu_{0,k}}$ and $\Pi^{\perp}_{\mb{A}_0} = \mb{I}_N - \mb{A}_0(\mb{A}^H_0\mb{A}_0)^{-1}\mb{A}_0^H$.

In \cite{For_EUSIPCO_19}, we showed that MUSIC functionals built upon robust $M$-estimators of the shape matrix, such us Tyler's and Huber's ones, are not efficient with respect to the SSCRB in \eqref{SSCRB}. Then, two questions arise now: can we obtain better performance by relying on the $R$-estimator $\widehat{\mb{V}}_{1,R}$ in \eqref{com_one_step_R}? In addition, will the resulting $R$-MUSIC functional $\hat{\bs{\nu}}_R = \bs{\Delta}_M(\widehat{\mb{V}}_{1,R})$ be semiparametric efficient with respect to the SSCRB in \eqref{SSCRB}? We will try to answer to these two important questions in the next Section.

\section{Numerical analysis}\label{numerical}
Finally, we present a numerical investigation of the $R$-MUSIC DOA estimator and of its semiparametric efficiency. Specifically, we compare its MSE on the estimation of the spatial frequency vector $\bs{\nu}_0$ with the SSCRB in \eqref{SSCRB}. We would like to underline however that our simulations have been performed in a realistic \virg{finite-sample} regime, i.e. for a reasonable number $L$ of snapshots, while, rigorously speaking, efficiency is an asymptotic ($L\rightarrow \infty$) property.

For simplicity, we consider a uniformly linear array (ULA) whose steering vector is given by $\mb{a}(\nu)=(1,e^{j2\pi\nu},\ldots,e^{j2\pi(N-1)\nu})^T$.\footnote{Let $d$ be the ULA element spacing and $\lambda$ its operating wavelength. Then $\nu = \nicefrac{d}{\lambda}\sin(\gamma)$ where $\gamma$ is the conic angle.} We simulate $L$ i.i.d. snapshots $\{\mb{z}_l\}_{l=1}^L$ according to the semiparametric CES model in \eqref{CES_model} under two different density generators leading to a set of 1) $t$-distributed data and of 2) Generalized Gaussian ($GG$) data.

The simulation parameters are set up as follows:
\begin{itemize}
	\item Two correlated sources ($K=2$) at spatial frequencies $\nu_1=0.1$ and $\nu_2=0.2$ with correlation matrix 
	\begin{equation*}
	\bs{\Gamma}_0 = \tonde{ \begin{array}{cc}\sigma_1^2 & \rho \sigma_1 \sigma_2\\ \rho \sigma_1 \sigma_2& \sigma_2^2 \end{array} }
	\end{equation*}
	where $\sigma_1^2 = \sigma_2^2  =  \sigma_0^2 \cdot 10^{(\mathrm{SNR}/10)}$,
	\item The noise power $\sigma_0^2 = 1$, the Signal-to-Noise ratio $\mathrm{SNR} = 5 \mathrm{dB}$ and the correlation coefficient $\rho = 0.5$,
	\item The number of snapshots is $L=5N$ where $N=8$,
	\item The number of Monte Carlo runs is $10^6$.
\end{itemize} 

For the purpose of comparison, we consider three MUSIC functionals:
\begin{itemize}
	\item \textit{SCM-MUSIC}: $\hat{\bs{\nu}}_{SCM} \triangleq \bs{\Delta}_M(\widehat{\mb{V}}_{1,SCM})$ where the Sample Covariance Matrix (SCM) is a standard estimator of the snapshot shape matrix and it is given by:
	\be
	\widehat{\mb{V}}_{1,SCM} = \frac{\widehat{\bs{\Sigma}}_{SCM}}{[\widehat{\bs{\Sigma}}_{SCM}]_{1,1}}, \quad \widehat{\bs{\Sigma}}_{SCM} = \frac{1}{L}\sum_{l=1}^{L} \mb{z}_l\mb{z}_l^H,
	\ee
	\item \textit{Tyler-MUSIC} \cite{Mahot}: $\hat{\bs{\nu}}_{Ty} \triangleq \bs{\Delta}_M(\widehat{\mb{V}}_{1,Ty})$, where the Tyler's estimator $\widehat{\mb{V}}_{1,Ty}$ has been showed in \eqref{Ty_schape},
	\item \textit{$R$-MUSIC}: $\hat{\bs{\nu}}_R = \bs{\Delta}_M(\widehat{\mb{V}}_{1,R})$ where the semiparametric efficient $R$-estimator $\widehat{\mb{V}}_{1,R}$ is given in \eqref{com_one_step_R}.  
\end{itemize}

As MSE indices, we adopt the following one:
\be
\varsigma_\alpha \triangleq E\{\norm{(\hat{\bs{\nu}}_\alpha - \bs{\nu}_0)(\hat{\bs{\nu}}_\alpha - \bs{\nu}_0)^T}_F\},
\ee
where $\norm{\cdot}_F$ is the Frobenius norm and $\alpha = \{SCM, Ty, R\}$. As SSCRB index, we report $\varepsilon_{SSCRB} = \norm{\mathrm{SSCRB}(\bs{\nu}_0|\bs{\zeta}_0,\sigma_0^2, h_0)}_F$. Note that the closed form expression of the SSCRB in \eqref{SSCRB} can be found in \cite{For_SCRB_complex,For_EUSIPCO_19}. 

\subsection{$t$-distributed data}
\begin{figure}[h]
	\centering
	\includegraphics[height=5.5cm]{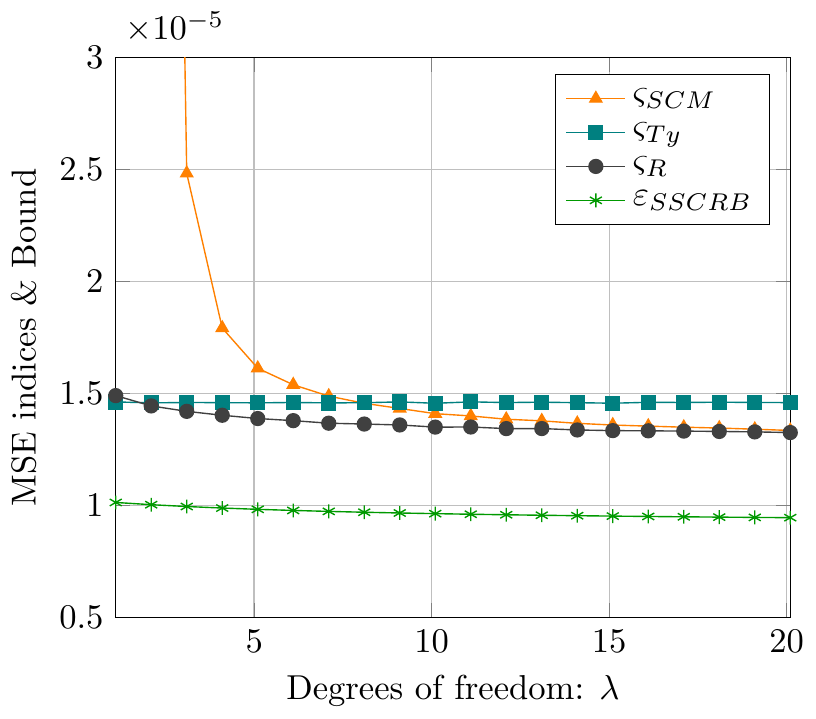}
	\caption{MSE indices and SSCRB vs $\lambda$ for $t$-distributed data.}
	\label{fig:Fig1}
\end{figure}

The complex $t$-distribution belongs to the CES family and its pdf can be obtained from the general expression in \eqref{true_CES} by using the following density generator \cite{Esa}: 
\be
\label{dg_t_dist}
h_0(t) = \frac{\Gamma(\lambda+N)}{\pi^{N}\Gamma({\lambda} )}\left( \frac{\lambda}{\eta}\right) ^{\lambda}\left( \frac{\lambda}{\eta} + t \right)^{-(\lambda+N)}, t \in \mathbb{R}^+,
\ee
where the degrees of freedom $\lambda \in (1,\infty)$ controls the data non-Gaussianity while the scale parameter $\eta$ has to be set as $\eta = \lambda/(\lambda-1)$ in order to satisfy the constraint in \eqref{const}. Fig. \ref{fig:Fig1} shows the MSE of the three considered MUSIC functionals as function of the non-Gaussianity parameter $\lambda$. As expected, for small value of $\lambda$, i.e.\ for highly  
non-Gaussian data, $\hat{\bs{\nu}}_{SCM}$ has very bad estimation performance due to the non robustness of the SCM estimator $\widehat{\mb{V}}_{1,SCM}$. On the other hand, for $\lambda \rightarrow \infty$ its performance improves since the data distribution collapses into the Gaussian one. Due to the well-known robustness properties of the Tyler's shape matrix estimator $\widehat{\mb{V}}_{1,Ty}$, the MSE of the Tyler-MUSIC functional $\hat{\bs{\nu}}_{Ty}$ is invariant with respect to $\lambda$ but its MSE index $\varsigma_{Ty}$ remains far from the SSCRB, in particular when the data tends to be Gaussian. Due to its semiparametric efficiency property, the $R$-MUSIC functional $\hat{\bs{\nu}}_{R}$, outperforms both the SCM-MUSIC and the Tyler-MUSIC for every value of the non-Gaussianity parameter $\lambda$. However, neither $\hat{\bs{\nu}}_{R}$ is able to achieve the SSCRB. Some hints about this lack of efficiency and some possible future research direction aiming at deriving semiparametric efficient DOA functionals will be provided in our conclusion collected in Sec. \ref{conclus}.

\subsection{$GG$-distributed data}
\begin{figure}[h]
	\centering
	\includegraphics[height=5.5cm]{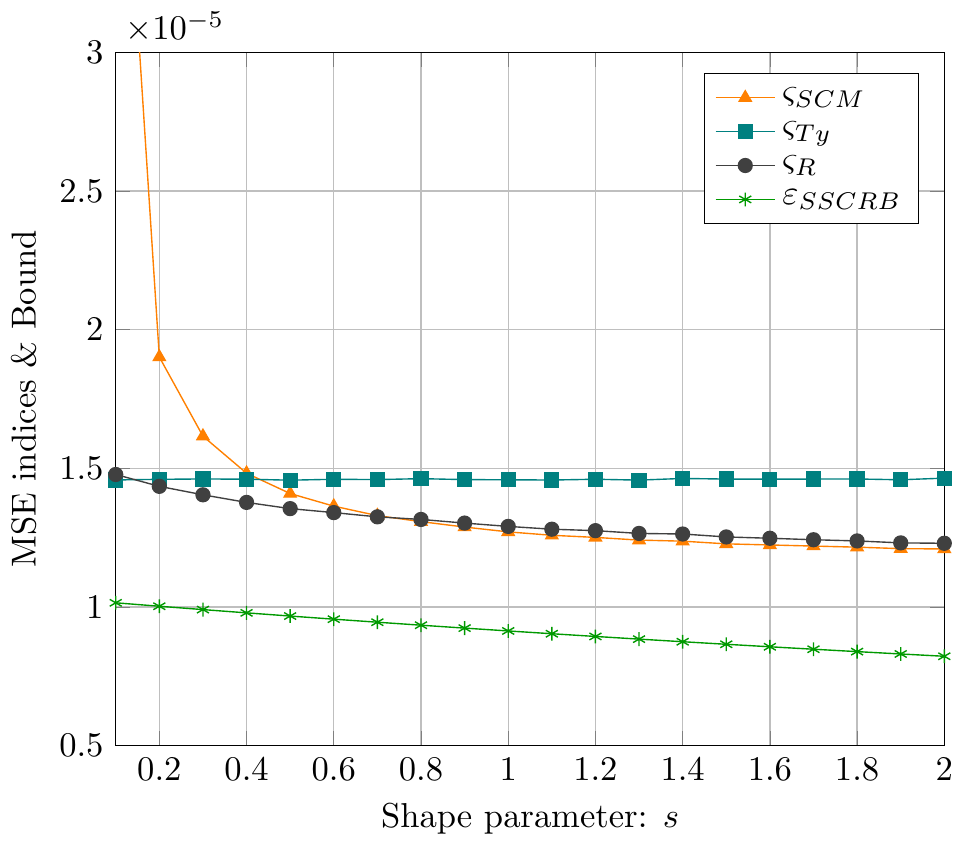}
	\caption{MSE indices and SSCRB vs $s$ for $GG$-distributed data.}
	\label{fig:Fig2}
\end{figure}

Another example of CES distribution is the Generalized Gaussian ($GG$) one. In particular, the $GG$ pdf can be obtained from the general expression in \eqref{true_CES} by using the following density generator \cite{Esa}:
\be
\label{dg_GG}
h_0(t) = \frac{s\Gamma(N)b^{-N/s}}{\pi^N\Gamma(N/s)}\exp\tonde{-\frac{t^2}{b}}, t \in \mathbb{R}^+.
\ee
The GG distribution is able to characterize data with heavier tails ($s<1$) and with lighter tails ($s>1$) as compared to a Gaussian dataset ($s=1$). The scale $b$ is a free parameter that, as for $\eta$ in the $t$-distribution, has to be set in order to satisfy the constraint is \eqref{const}. In particular, we have that $b=\quadre{N\Gamma(N/s)/\Gamma((N+1)/s)}^s$. On the same line of Fig. \ref{fig:Fig1}, Fig. \ref{fig:Fig2} shows the MSE indices of the tree MUSIC functional as function of the non-Gaussianity parameter $s$ for GG-distributed data. The simulation results in Fig. \ref{fig:Fig2} confirm the ones previously discussed for $t$-distributed data:

\begin{itemize}
	\item The MSE of the Tyler-MUSIC functional $\hat{\bs{\nu}}_{Ty}$ is invariant with respect to the data non-Gaussianity. However, its MSE index is far from the SSCRB, in particular for Gaussian ($s=1$) and sub-Gaussian ($s>1$) data.
	\item The SCM-MUSIC functional $\hat{\bs{\nu}}_{SCM}$ suffers in heavy tailed data ($s<1$), while outperforms the Tyler-MUSIC in Gaussian ($s=1$) and sub-Gaussian data ($s>1$).
	\item The $R$-MUSIC functional $\hat{\bs{\nu}}_{R}$ outperforms both Tyler-MUSIC and SCM-MUSIC estimators in heavy-tailed scenarios $(s<1)$, while its MSE is comparable to the one of the SCM-MUSIC functional in Gaussian ($s=1$) and sub-Gaussian ($s>1$) environment.
\end{itemize}

This simulative investigation of the performance of the three MUSIC estimation functionals leaves open a fundamental question: is it possible to derive a semiparametric efficient estimator of source spatial frequencies for CES distributed snapshots? A preliminary answer and some related hints will be given in our concluding remarks.
  
\section{Concluding remarks}\label{conclus}
The goal of this paper was twofold: following \cite{visuri, Esa_DOA, abeida_SP_paper}, we first aimed at reformulating the classical Gaussian-based snapshot model for DOA estimation in a much more general and realistic semiparametric CES model. Then, the semiparametric efficiency of a MUSIC DOA functional exploiting the recently derived shape matrix $R$-estimator \cite{Sem_eff_est_TSP} has been assessed through numerical simulations. This preliminary investigation has shown that, in heavy-tailed data, the $R$-MUSIC functional outperforms (at least) two classical MUSIC estimators built from the SCM and the Tyler's $M$-estimator of shape. However, it fails to be semiparametric efficient with respect to the SSCRB \cite{For_SCRB_complex,For_EUSIPCO_19}. We believe that one of the reason of this lack of efficiency is in the fact that the MUSIC functional is based on the eigen-decomposition of the (estimated) snapshot shape matrix. Consequently, even if the adopted shape matrix estimator is semiparametric efficient (as the $R$-estimator in \cite{Hallin_Annals_Stat_2, Sem_eff_est_TSP}), this does not imply that its eigenvectors, obtained for example through a singular value decomposition (SVD), represent a semiparametric efficient estimate of the true eigenspace. This means that, instead of implementing an estimator of a shape matrix and, from it, evaluate the eigenvectors to be used in the MUSIC functional, we should estimate the noise eigenspace \textit{directly} from the collected snapshots. To this end, the recent work \cite{Hallin_PCA} has investigated the possibility to derive \textit{distributionally robust} and \textit{semiparametric efficient} estimators of the eigenvectors of the shape matrix of a CES distributed dataset. This promising approach could be the key to obtain semiparametric efficient sub-space-based DOA estimators in general non-Gaussian and heavy-tailed environment. In addition to this, future works will aim at providing a performance comparison with other existing robust DOA estimation methods. Among others, the comparison with the $G$-MUSIC algorithm \cite{Romain} is of both practical and theoretical importance. In fact, building upon robust statistics and random matrix theory (RMT), the $G$-MUSIC DOA estimator has been proved to have better performance with respect to others robust competitors. However, the study of its semiparametric efficiency is still an open problem.         
 
\bibliographystyle{IEEEtran}
\bibliography{RadConf_ref_semipar_eff_estim}

\end{document}